\documentclass[12pt,letterpaper]{article}
\usepackage{array}
\usepackage{geometry}
\usepackage{microtype}
\usepackage{setspace}
\usepackage{lscape}

\begin{document}

~

\noindent {\Large Co-Occurring HIV Risk Behaviors among Males
  Entering Jail} \date{}

~

\subsubsection{Authors:} 
Tao Liu, PhD$^1$ (tliu@stat.brown.edu)\\
Lauri Bazerman, MS$^{2,3}$ (lbazerman@gmail.com)\\
Megan Pinkston, MD$^{2,4}$ (mpinkston@lifespan.org)\\
Amy Nunn, PhD$^{2,3,5}$ (amy\_nunn@brown.edu),\\
Aadia Rana, MD$^{2,5}$ (arana@lifespan.org)\\
Curt G. Beckwith, MD$^{2,3,5}$ (cbeckwith@lifespan.org)

\subsubsection{Institutional Affiliations:}
$^1$Department of Biostatistics, Center for Statistical Sciences, Brown University School of Public Health, Providence, RI; $^2$The Miriam Hospital, Providence, RI; $3$ The Lifespan/Brown Center for Prisoner Health and Human Rights; $^4$Department of Psychiatry and Behavioral Medicine, the Alpert Medical School of Brown University, Providence, RI; $^5$Department of Medicine, the Alpert Medical School of Brown University, Providence, RI. 

\subsubsection{Corresponding Author:} 
Tao Liu, PhD, Associate Professor of Biostatistics;\\ Address: Brown University, Box GS121-7, Providence, RI 02912, U.S.;\\ Telephone Number: 401-863-6480; \\Fax Number: 401-863-9182.

\subsubsection{Sources of Funding Support:}
The research is supported by the Providence/Boston Center for AIDS Research (grant P30AI42853).  Dr. Pinkston's work is partially supported by a National Institute of Mental Health grant (5R01MH084757). 

\subsubsection{Conflicts of Interest:}
The authors declare that there is no conflict of interest regarding the publication of this paper.

\pagebreak 

\begin{center}
{\large \bf Abstract}
\end{center}

People going through the United States (US) criminal justice system 
often exhibit multiple behaviors that increase their risk of HIV infection and 
transmission.  This paper examined the pattern of co-occurring HIV risk 
behaviors among male jail detainees in the US.  We conducted multivariate 
analyses of baseline data from an HIV intervention study of ours, and found 
that: (1) cocaine use, heroin use and multiple sexual partners; and (2) heavy 
drinking and
marijuana were often co-occurring among this population.  From pairwise
analyses, we also found that (1) heroin and IDU (2) unprotected sexes with
main, with non-main, and in last sexual encounter were mostly
co-occurring behaviors. Further analyses of risk behaviors and demographic 
characteristics of the population showed that IDU were more prevalent among 
middle ages (30-40)
and multiple prior incarcerations, and having multiple sex partners was more 
prevalent among
young males younger than 30 years, African American race, and those with low
education.  Our findings suggest that efficient interventions to
reduce HIV infection in this high-risk population may have to target
on these behaviors simultaneously and be demographically adapted.

~

\noindent \textbf{Keywords}: HIV risk, co-occurring behaviors,
correctional facilities, male jail detainees.  

\pagebreak

\section{Introduction}
\label{sec:introduction}

Over seven million people passed  through the criminal
justice system in the United State (US) in year 2012 \cite{Glaze2012}.  Among 
this population, it was estimated that about 2\% was infected with HIV including 
those unaware of   their   infection  \cite{CDC,Spaulding2002,Spaulding2009} --- 
as a contrast, the prevalence among the US adult population is around 0.3\% 
according to the US Centers for Disease Control and Prevention (CDC).  The
prevalence of HIV  infection within jails and prisons was estimated to be about 
3 to 6
times  higher  compared with that  among  non-incarcerated  populations
\cite{Spaulding2009,2002,Dean-Gaitor1999,Hammett1998,Maruschak2006}.

The  reasons for  this increased  burden of  HIV among  populations in
correctional settings are multi-factorial  and include increased rates
of  substance abuse,  mental illness,  poverty and  health disparities
\cite{Springer2004}.  Persons  who interact with the  criminal justice
system may be  disenfranchised from health services  in the community,
such as screening programs. That makes  the time of incarceration an important 
public
health opportunity to provide HIV  prevention and testing services and
linkage to care \cite{Harrison2006,Braithwaite1996,Braithwaite2008}.

The time period preceding  incarceration has been shown to be characterized by 
increased substance use  and risky  sexual behaviors  that increased  exposure 
to
HIV,    viral    hepatitis,    and    other    transmitted    diseases
\cite{Wohl2000,Conklin2000,Altice1998,Stephenson2006,Weinbaum2005,Mertz2002}.
Release from correctional  facilities might also be a  time of high-risk
of  acquiring  or  spreading  infections  as  persons  re-entered  their
communities         and          resumed         risk         behaviors
\cite{Chandler2009,Morrow2009,Milloy2009}.  Thus, correctional-based   HIV
counseling and testing programs  and prevention interventions may help
to  decrease   their  risk   behaviors  following  release   from  the
correctional environment  and therefore  reduce new HIV  infections in
this as well as the general population.

Although studies  have documented prevalent (direct  and indirect) HIV
risk  behaviors  before  entering   jail  (including  heavy  drinking,
substance   abuse,   sexual    promiscuity,   and   unprotected   sex)
\cite{Morrow2009,Milloy2009,McCoy2004,Shannon2008,WHO2007,Fazel2006,koulierakis2002,Chandler2009}
there is limited understanding of the interrelationships among these
risk  factors.   To  effectively target  prevention  interventions  to
persons at the  greatest risk of HIV infection  among this population,
it  is critically  important  to understand  their  risk profiles  and
quantify which  risk behaviors are  more likely to co-occur.   In this
paper,  co-occurring behaviors  are  defined as  behaviors that  occur
within  certain  time  period  (e.g.\  a  3  months  window)  and  not
necessarily  always in  the same  episode (i.e.\  concurrently).  This
definition is  consistent with  the need  of broader  interventions on
behaviors that  are predictive  of (not necessarily  determinative of)
each other  and jointly place  an individual at  a higher risk  of HIV
infection.

In this paper, we conducted a secondary analysis of data from a study on
HIV counseling and testing in jail \cite{Beckwith2009}.  Specifically,
we  used the  baseline data  of the  study to  investigate: 1)  whether
certain risk  behaviors were co-occurring and to  what extent,  and 2)
whether  risk  behaviors were prevalent among people with  certain  demographic
characteristics.

\section{Methods}
\label{sec:methods}

\subsection{Prior Study and Data}
\label{sec:data}

We previously conducted a two-arm randomized study \cite{Beckwith2009}
to assess  HIV risk  behaviors among males  entering the  Rhode Island
Department of  Corrections (RIDOC) jail  and compared the  efficacy of
two methods of  HIV counseling and testing  (conventional versus rapid
HIV testing) with respect to reducing post-release HIV risk behaviors.
A total of  264 HIV-negative males met the  study enrollment criteria,
provided the written informed consent,  were recruited within 48 hours
of incarceration, and completed the  study.  The study was approved by
the  Miriam  Hospital institutional  review  board,  the Rhode  Island
Department of Corrections (RIDOC) Medical Research Advisory Group, and
the Office for Human Research  Protections of the Department of Health
and Human Services.  More details of the study are available elsewhere 
\cite{Beckwith2009}.

In this paper, we focused on data that were collected at the baseline of the 
study, including demographic information and self-reported  HIV risk behaviors 
during 3 months prior  to incarceration.  The self-reported  risk behaviors were 
collected using a written quantitative behavioral assessment survey on
participant's recent  drinking, substance use behaviors  (cocaine use,
heroin use, marijuana use, injection of any drug) and sexual behaviors
(multiple sexual  partners, unprotected sex at  last sexual encounter,
unprotected sex with  main partner, and unprotected  sex with non-main
partner).  Because  only data  at the  baseline prior  to intervention
randomization were used  in this  paper, we  did not  distinguish study
participants by their study arms.

\subsection{Statistical Analyses}

We  conducted  three sets of statistical  analyses, outlined as follows:

\emph{Analysis I: } The co-occurrence of two risk behaviors (pair-wise
analysis) was  assessed using  logistic regressions where  one behavior
(Behavior 1) was used as the dependent variable, and the other behavior
(Behavior  2) as  an predictor  variable.   The results  are shown  in
Table~\ref{tba:pair-wise}.   All  regressions were  adjusted  for  the
following demographic covariates: age (categorized as $< 25$; $25 \sim
30$;  $30  \sim  40$;  and  $> 40$  years),  race  (Caucasian;  Black;
Hispanic;  others), number  of prior  incarcerations (dichotomized  at
median: $< 7$; $\ge 7$), length  of incarceration as severity index of
crime leading to incarceration ($< 2$ wks; 2 wks $\sim 1/2$ yr; $>1/2$
yr), and education (did not finish high school; otherwise).

Pair-wise co-occurring risk behaviors were quantified using odds ratios
(ORs), where an OR $> 1$ (OR  $<1$) suggests that the existence of one
behavior was predictive  of the  existence (or  absence) of  the other
behavior.

We used  the available complete  data for assessing risk  behaviors, so
the analysis sample  size varied (range: 73-256 as in  Tables 1 \& 2).
The overall missing data on risk  behaviors were moderate ($< 5\%$), if
we  did  not count  systematic  missingness  as missing  values  (e.g.\
Missing   sexual  behaviors   for  those   without  sexual   partner).
Throughout,   we  made   the  missing   at  random   (MAR)  assumption
\cite{Rubin1976}; that is,  we assumed that with  the same demographic
profile,  those   who  provided  complete  answers   to  the  baseline
questionnaire had engaged  in similar risk behaviors as  those who did
not \cite{Little2002}.

\emph{Analysis II:} Multiple co-occurring  risk behaviors were assessed
using  multivariate  logistic  regressions  where  one  risk  behavior
(Behavior 1)  was used  as the dependent  variable and  other behaviors
(Behaviors  2)   as  predictor  variables. The results are shown in  
Table~\ref{tba:all}.
Similar to  Analysis I,  the co-occurring of  Behavior 1  with other
behaviors was characterized by  ORs, which have similar interpretations
except  that  the  ORs   in  Table~\ref{tba:all}  are  conditional ORs after 
accounting  for  all  other  behaviors of  (Behaviors  2).   Again, all analyses 
were adjusted for the same set of demographic characteristics
as in Analysis I.  When heavy drinking, cocaine, heroin, marijuana and
multiple sexual partners were used as dependent variables, we excluded
the  risky   behaviors  `unprotected   sex  with  main   partner'  and
`unprotected sex  with non-main partner',  because they only  applied to
subsets of study participants with  sexual partners and including them
would reduce  the sample size by  half and reduce the  analysis power.
When `unprotected  sex with  main partner'  and `unprotected  sex with
non-main partner'  were used  as the dependent  variables, `unprotected
sex at  last sexual  encounter' was  excluded from  predictor variables
because  the later  behavior  strongly correlated  with the former behaviors and  
therefore
overwhelmed the associations of the former two behaviors with other risk
factors.  IDU was excluded from  the analysis because the prevalence of
injection drug  use was low  (overall 8\%)  leading to sparse  data for
multivariate analysis  and unreliable  estimates due to  collinearity of
heroin use and IDU \cite{Armitage2002}.

\emph{Analysis III:} Further, we  examined the associations between HIV
risk behaviors and various  demographic characteristics using logistic
regressions, where each risk behavior was used a dependent variable and
predictor  variables   included:  age,  race,  the   number  of  prior
incarcerations,  length of  incarceration as  severity index  of crime
leading to incarceration, and  education.  The predictor variables were
categorized in the same way as in Analyses I and II.  The associations
of  each   risk  behaviors   and  certain  demographic   profiles  were
characterized by ORs.

Data were  extracted and  prepared using Access  2003 \cite{microsoft}.
All   analyses  were   conducted  using   the  statistical   program  R
\cite{R-Project2012}.   Analysis   lack  of  fit  was  assessed  using
Hosmer-Lemeshow  tests.    Statistical  significance was  set   at  a
\emph{p}-value $< 0.05$.

\section{Results}
\label{sec:results}

Among the  264 male HIV-negative  participants, the median age  was 30
years (range 18-65); the majority  was Caucasian (52\% Caucasian, 22\%
Black, 14\% Hispanic,  12\% others); 51\% did not  finish high school;
and the median number of  lifetime incarcerations was 6 (range 1-200).
Within the  prior 3 months  before incarceration, 103 (39\%,  data not
available  (NA) =  1) were  heavy  drinkers; 27  (10\%, NA  = 1)  used
heroin; 100  (38\%, NA  = 1)  used cocaine;  161 (61\%,  NA =  1) used
marijuana; and 22  (8\%, NA = 1)  had injected any type  of drug.  For
the same  time period,  203 (77\%)  had a main  sexual partner  and of
those  170 (84\%,  NA =  1)  never used  a  condom; 111  (42\%) had  a
non-main sexual partner  and of those 37  (36\%, NA = 3)  never used a
condom;  81 (31\%)  had both  main  and non-main  sexual partners;  61
(26\%, NA = 4) had multiple  ($\ge 3$) recent sexual partners; and 233
(90\%, NA = 4) did not use a condom at last sexual encounter.

In Analysis I, cocaine use was found to be highly predictive of heroin  use (OR 
= 5.21 with a  95\% confidence interval
(CI) of 1.8 - 15), IDU (OR = 6.65, CI = 1.9 - 23), and multiple sexual
partners (OR =  2.45, CI = 1.1 -  5.3); see Table~\ref{tba:pair-wise}.
Heroin use and IDU were mostly co-occurring, suggesting that injection
might  be the  preferred  route  of heroin  use.   Heavy drinking  and
marijuana use were predictive  of each other  (OR = 2.88,  CI =  1.6 -
5.3).   Participants who  had  unprotected sex  with  their main  and
non-main sexual partner(s) were more  likely to have unprotected sex at
last sexual encounter (OR = 25.7, CI =  9.1 - 73.0 and OR = 88.6, CI =
15 - 200,  respectively).  Notably, unprotected sex  with main partner
and with  non-main partner(s) was likely  to co-occur (OR =  6.43, CI =
1.56  - 78.8).   In terms  of protective  behaviors, participants  who
reported IDU  and those with  multiple sexual  partners were found  to be
more  likely to  use condoms  at  ``the last sexual  encounter'', though  this
finding was marginally statistically insignificant (\emph{p}-values  = 0.08 and
0.07, respectively).

In Analysis II, we found  that (1) cocaine use, heroin use, and multiple  sexual 
partners, and (2) heavy drinking
and  marijuana  use were mostly co-occurring  (Table~\ref{tba:all}).   Heavy
drinking and marijuana  use were highly predictive of each  other (OR =
3.40, CI  = 1.7 -  7.1).  Cocaine use was predictive of heroin use  (OR =
9.20, CI = 2.7 - 38.7)  and multiple ($\ge 3$) sexual partnerships (OR
= 2.56; CI = 1.1 - 6.0).

The analyses that examined the relationships between risk behaviors and
demographic characteristics (Analysis III)  showed that male  jail detainees
with age between 30  - 40 were more likely to abuse  cocaine (OR = 8.6,
CI = 3.5  - 23.2), heroin (OR = 4.7,  CI = 1.2 - 23.7), and  IDU (OR =
2.8, CI = 1.2 - 6.9).  Younger males  with age $<$ 30 were more likely to
abuse marijuana (OR  = 3.8, CI =  2.2 - 6.9) and  had multiple sexual
partners (OR = 2.1, CI = 1.2 - 3.8).  African American were more likely
to have multiple sexual partners (OR =  3.7, CI = 1.8 - 7.9), but less
likely to  engage in unprotected  sex in  last sexual encounter  (OR =
0.3, CI = 0.1 - 0.6), with main partner (OR = 0.3, CI = 0.1 - 0.9) and
non-main partner(s) (OR =  0.1, CI = 0.03 - 0.4).   Having more than 7
prior incarcerations was predictive of heavy drinking (OR = 1.8, CI =
1.1 - 3.2), cocaine use (OR = 2.5, CI = 1.4 - 4.6), and IDU (OR = 2.9,
CI = 1.1 - 7.7).  Finishing high school was predictive of having less sexual
partners (OR  = 0.5,  CI =  0.3 - 0.9)  but more likely engaging in
unprotected sex in  last sexual encounter (OR  = 3.3, CI =  1.6 - 7.2)
and with main sexual partner (OR = 3.2, CI = 1.3 - 8.0).

\section{Discussion}
\label{sec:discussion}

Our  results indicate  that males  entering  jail exhibit high rates  of
substance use  and sexual risk  behaviors that increase their  risk of
HIV  and  other  infectious  diseases.  Our  study  adds  to the existing
literature  by demonstrating  high risk  behaviors among  incarcerated
populations  and by  highlighting whether  certain risk  behaviors are
more  likely  to  be  co-occurring   thus  compounding  risk  for  HIV
infection.

Particularly from our pairwise and multivariate analyses, we find that
cocaine is  co-occurring with  several other risk  behaviors including
heroin use, injection drug use, and multiple sexual partners.  Cocaine
use has  been reported  to not  only increase  the probability  of HIV
transmission, but also the potential  of poor health outcomes in those
living                with                HIV                infection
\cite{McCoy2004,Strathdee2003,Moore2001,Shannon2008,Lucas2006}.  Given
that  there is  currently  no pharmacotherapy  based intervention  for
cocaine addiction as there is for opiate addiction, our study supports
the need of developing  behavior-based interventions for cocaine abuse
that  is  appropriate  for  incarcerated populations  in  addition  to
addressing  opiate  use  and   risky  sexual  behaviors.   Since  jail
incarcerations   may  be   as  short   as  several   days,  behavioral
interventions     such     as      contingency     management     (CM)
\cite{Petry2007,Rash2008,Ledgerwood2008,Petry2012c}     may    provide
immediate reinforcement for abstinence from cocaine use, and cognitive
behavioral  interventions that  are paired  with CM  upon release  may
offer a  bridge for continued abstinence  following community re-entry
\cite{Epstein2003}.   However,  these   interventions  have  not  been
implemented among incarcerated populations \cite{Polonsky1994}.

The finding  that unprotected  sex with  main partner  is co-occurring
with  unprotected sex  with non-main  partner(s) is  another important
finding, as  this suggests  that some  participants could  be involved
with concurrent sexual  relationships.  Concurrent sexual partnerships
in  incarcerated populations  have  been reported  in several  studies
\cite{Adimora2004,Mumola2006,Adimora2003,Khan2009,Manhart2002}.
Further   accounting   for   concurrent   sexual   partnerships   (and
social/sexual  networks)   in  our   analyses  would   strengthen  our
conclusions,  but  unfortunately  as  one limitation  of  this  paper,
collecting concurrent  behaviors data is  not a focus of  our original
study.

Heavy alcohol use and marijuana use  are common substances used by this
population  and found  to be  mostly co-occurring.   Previous findings
with younger incarcerated men  \cite{Valera:2009} suggested that prior
to  incarceration,  the  use  of marijuana  alone  and  alcohol  alone
increased the likelihood of multiple sexual partners (i.e.\ 3 or more)
and  when  in  used  in combination,  sexual  HIV-risk  behaviors  and
inconsistent  condom use  behaviors  with  female partners  increased.
Similar  finding also  can  be  found in  \cite{Edlin1994,Altice1998}.
Comparable  to other  drugs of  abuse, alcohol  and marijuana  use can
impair  judgment thereby  preventing  safer sex  behaviors, and  hence
remain an important domain for intervention.

This  paper has  several  limitations.  The  risk  behavior data  were
self-reported  which  might  have  introduced  bias  and  possibly  an
underreporting  of  risk  behaviors  given the  environment  in  which
participants  completed  the  questionnaire.   Our  findings  are  not
generalizable to incarcerated women  or the entire population, because
it  is known  that incarcerated  women have  a different  rate of  HIV
infection and other transmissible diseases compared to men.  The study
sample size is  limited and study participants are  restricted only to
those at  the RIDOC, which limits  our analysis power to  identify all
co-occurring risk behaviors.

As  the   U.S.\  incarceration   population  continues  to   grow  and
disproportionate  rates  of  HIV  infection  continue  to  rise  among
incarcerated  individuals,  the   implications  for  intervention  are
important  and imperative.   Jails  provide a  unique opportunity  for
structural interventions  for this high-risk population.   The results
of this  study offer  more insight  into the  risk behaviors  of males
entering the  RIDOC jail,  and elucidate the  educational, counseling,
and intervention  needs of men  at risk  for HIV infection  within the
criminal justice system.


\begin{landscape}
\begin{table}[!p]
  \caption{Pair-wise association among risk behaviors.} 
  \label{tba:pair-wise}
  {\footnotesize
  \begin{tabular}{p{4cm}| p{1.6cm} p{1.3cm} p{1.2cm} p{1.3cm} p{1.2cm} p{1.5cm} p{1.6cm}  p{1.6cm}  p{1.6cm}}
    \hline
    & \multicolumn{9}{c}{\textbf{Behavior 2}}  \\
    \cline{2-10}
    \textbf{Behavior 1} & Heavy drinking & Cocaine & Heroin & Marijuana & Injection drug & Multiple sexual partners & Unprotected sex at last sexual encounter &Unprotected sex with main partner  &Unprotected sex with non-main partner(s) \\
    \hline
Heavy drinking & --- & 1.32 (n=256) & \emph{0.42} (n=256) & \textbf{2.88} (n=256) & 0.45 (n=256) & 1.28 (n=253) & 1.06 (n=256) & 0.62 (n=196) & 0.93 (n=102) \\
\hline
Cocaine & 1.34 & --- & \textbf{5.21} (n=257) & 1.29 (n=257) & \textbf{6.65} (n=257) & \textbf{2.45} (n=253) & 0.74 (n=227) & 0.69 (n=197) & 1.00 (n=102) \\
\hline
Heroin & \emph{0.45} & \textbf{5.22} & --- & 0.54 (n=257) & $>$\textbf{30} (n=257) & 1.52 (n=253) & 0.53 (n=227) & 0.56 (n=197) & 0.51 (n=102) \\
\hline
Marijuana & \textbf{2.89} & 1.30 & 0.60 & --- & 0.56 (n=257) & \emph{1.94} (n=253) & 0.52 (n=227) & 0.67 (n=197) & 0.63 (n=102) \\
\hline
Injection drug & 0.47 & \textbf{6.64} & $>$\textbf{30} & \emph{0.51} & --- & 2.04 (n=253) & \emph{0.36} (n=227) & 0.65 (n=197) & 0.63 (n=102) \\
\hline
Multiple sexual partners & 1.39 & \textbf{2.46} & 1.42 & \emph{2.00} & 1.89 & --- & \emph{0.50} (n=227) & 1.02 (n=197) & \emph{0.44} (n=102) \\
\hline
Unprotected sex at last sexual encounter & 1.04 & 0.74 & 0.46 & 0.54 & \emph{0.32} & \emph{0.51} & --- & \textbf{25.7} (n=197) & $>$\textbf{30} (n=102) \\
\hline
Unprotected sex with main partner & 0.61 & 0.68 & 0.49 & 0.65 & 0.61 & 1.00 & \textbf{24.4} & --- & \textbf{6.43} (n=73) \\
\hline
Unprotected sex with non-main partner(s) & 0.95 & 0.95 & 0.47 & 0.66 & 0.49 & \emph{0.44} & $>$\textbf{30} & \textbf{8.62} & --- \\
 &  &  &  &  &  &  &  &  &  \\
\hline
\end{tabular}\\
(a) The table is not symmetric because the analyses are adjusted for
  the following covariates as predictors of Behavior 1: age, race, prior
  incarcerations, length of incarceration, and education.  \\
(b) The numbers in parentheses are sample sizes. \\
(c) \textbf{Bold} indicates a $p$-value $<0.05$ and \emph{italic} $<0.10$. 
}
\end{table}
\end{landscape}

\begin{table}[!p]
  \caption{Multivariate analyses of co-occurring risk behaviors}
  \label{tba:all}
  {\footnotesize
  \begin{tabular}{p{4cm}| p{1.3cm} p{1.3cm} p{1.3cm} p{1.6cm} p{1.6cm} p{2.2cm}}
    \hline
    &\multicolumn{6}{c}{\textbf{Behaviors 2}} \  \\
    \cline{2-7}
    \textbf{Behavior 1} & Heavy drinking & Cocaine & Heroin &
    Marijuana & Multiple sexual partners & Unprotected sex at
    last sexual encounter \\
    \hline
Heavy drinking (n=226) & --- & 1.74 & 0.55 & \textbf{3.40} & 1.23 & 1.21 \\
\hline
Cocaine (n=226) & 1.68 & --- & \textbf{9.20} & 0.91 & \textbf{2.56} & 0.95 \\
\hline
Heroin (n=226) & 0.41 & \textbf{10.2} & --- & 0.54 & 0.91 & 0.39 \\
\hline
Marijuana (n=226) & \textbf{3.30} & 0.95 & 0.46 & --- & 1.84 & 0.57 \\
\hline
Multiple sexual partners (n=226) & 1.11 & \textbf{2.38} & 1.06 & 1.91 & --- & 0.56 \\
\hline
Unprotected sex at last sexual encounter (n=226) & 1.26 & 0.91 & 0.54 & 0.51 & 0.55 & --- \\
\hline
Unprotected sex with main partner (n=196) & 0.66 & 0.80 & 0.56 & 0.72 & 1.15 & *** \\
\hline
Unprotected sex with non-main partner(s) (n=102) & 0.90 & 1.23 & 0.43 & 0.63 & \emph{0.44} & **** \\
\hline
\end{tabular} \\
(a) The table is not symmetric because the analyses are adjusted for the
  following demographic covariates: age, race, prior
  incarcerations, length of incarceration, and education.  \\
(b) The numbers in parentheses are sample sizes. \\ 
(c) \textbf{Bold} indicates a $p$-value $<0.05$ and \emph{italic}
  $<0.10$. \\ 
(d) *** ``Unprotected sex at last sexual encounter'' is
  not included as a predictor variable in the model.
} 
\end{table}

\end{document}